\documentstyle[aps,preprint,pra]{revtex}
\tighten
\renewcommand{\epsilon}{\varepsilon}

\begin{document}

\draft
\title{Efficient Quantum Key Distribution}
\author{M. Ardehali,$^1$
H. F. Chau,$^2$\footnote{e-mail: hfchau@hkusua.hku.hk} and Hoi-Kwong
 Lo$^3$\footnote{e-mail: hkl@hplb.hpl.hp.com}}
\address{$^1$ Research Labs, NEC Corporation, Sagamihara, Kanagawa 229, Japan
and 4-31-12-302, Atago Center, Tama-shi, Tokyo, Japan}
\address{$^2$ Department of Physics, University of Hong Kong, Pokfulam Road,
 Hong Kong}
\address{$^3$ Hewlett-Packard Labs, Filton Road, Stoke Gifford, Bristol
 BS12 6QZ, United Kingdom}
\date{\today}
\maketitle
\begin{abstract}
We devise a simple modification that essentially doubles the
efficiency of a well-known quantum key distribution scheme proposed
by Bennett and Brassard (BB84). Our
scheme assigns significantly different probabilities
for the different polarization bases
during both
transmission and reception to reduce the fraction of
discarded data. The actual probabilities used in the scheme are announced
in public. As the number of transmitted signals increases,
the efficiency of our scheme can be made to approach 100\%.
An eavesdropper may try to break such a scheme by
eavesdropping mainly along the predominant basis.
To defeat such an attack,
we perform a refined analysis of accepted data:
Instead of lumping all the accepted data together
to estimate a single error rate, we separate the accepted data
into various subsets according to the basis employed
and estimate an error rate for each subset {\it individually}.

\end{abstract}

\medskip
\pacs{\noindent
 \begin{minipage}[t]{5in}
 \end{minipage}
}

\section{Introduction}
\label{S:Intro}
As an encryption scheme is only as secure as its key,
key distribution is a big problem in
conventional cryptography.
Public-key based key distribution
schemes such as the Diffie-Hellman scheme~\cite{DH}
solve the key distribution problem by
making computational assumptions such as that
the discrete logarithm problem
is hard. However, unexpected future advances in algorithms and
hardware (e.g., the construction of a quantum computer
\cite{Shor94,Shor95}) may render
public-key based schemes insecure. Worse still, this would lead
to a {\it retroactive\/} total security break with disastrous consequences.
A big problem in conventional public-key cryptography is that there is, in
principle, nothing to
prevent an eavesdropper with infinite computing power
from passively monitoring the key distribution
channel and thus successfully decoding any subsequent communication.

Recently, there has been much interest in using quantum mechanics
in cryptography. [The subject of quantum cryptography was
started by S. Wiesner \cite{Wiesner} in a paper that was written
in about 1970 but remained unpublished until 1983. For a review
on the subject, see
\cite{sciam}.]
The aim of quantum cryptography has always been to solve problems
that are impossible from the perspective of conventional cryptography.
This paper deals with quantum key distribution \cite{BB84,Mor,Ekert} whose
goal is to detect eavesdropping using the laws of
physics. [Another class of applications of
quantum cryptography has also been proposed\cite{BBCS,BCJL}.
Those applications are mainly
based on quantum bit commitment.
However, it is now known \cite{LoChau1,Mayers2} that
unconditionally secure quantum bit commitment is
impossible. Furthermore, some other quantum cryptographic schemes
such as quantum one-out-of-two oblivious transfer have
also been shown to be insecure \cite{Lo}. For a review, see \cite{special}.]
In quantum mechanics, measurement is not just a passive, external process,
but an integral part of the formalism. Indeed, passive
monitoring of transmitted signals is strictly forbidden in quantum
mechanics. The quantum no-cloning \cite{Dieks82,WZ82}
theorem dictates that the copying of an unknown quantum state
would violate linearity, unitarity and causality---three cherished basic
physical principles in quantum mechanics.
Moreover, an eavesdropper who is listening to a channel in an attempt
to learn information about quantum states will almost always introduce
disturbance in the transmitted quantum signals~\cite{BBM}.
Such disturbance can be detected with high probability
by the legitimate users. Alice and Bob will use the transmitted
signals as a key for subsequent communications only when the
security of quantum signals is established
(from the low value of error rate).

Various quantum key distribution schemes have been proposed.
To illustrate our main ideas, we will use
the most well-known quantum key distribution
scheme (BB84) proposed by Bennett and Brassard \cite{BB84} in 1984.
The details of BB84 will be discussed in the next section. Here
it suffices to note two of its characteristics. Firstly,
in BB84 each of the two users, Alice and Bob, chooses for each
photon between
two polarization bases {\it randomly\/} (i.e.,
with equal probability) and independently. For this reason,
half of the times they are using different basis, in which case
the data are rejected immediately. Consequently, the efficiency of
BB84 is at most 50\%. Secondly, a naive error analysis is performed
in BB84. All the accepted data (those that are encoded and
decoded in the same basis) are lumped together and a {\it single\/} error
rate is computed.

In contrast, in our present scheme each of Alice and Bob
chooses between the two bases independently but
with {\it substantially different\/} probabilities.
As Alice and Bob are now much more likely to be using the same
basis, the fraction of discarded data is greatly reduced,
thus achieving a significant gain in efficiency.

An eavesdropper may try to break this new scheme by eavesdropping
mainly along the the predominant basis. To foil this attack,
a refined error analysis is performed. The accepted data
are further divided into two subsets according to
the actual basis used by Alice and Bob and
the error rate of each subset is computed {\it separately}.
We will argue that such a refined error analysis is necessary and
sufficient in ensuring the security of our improved scheme
against such a biased eavesdropping attack.

Our scheme is worth studying for several reasons.
Firstly, up till now, only a few quantum key distribution schemes have been
proposed.
The construction of any interesting new scheme is, therefore, a major
achievement in itself. Secondly, none of the existing schemes
based on non-orthogonal quantum cryptography has an
efficiency more than $50\%$.
[The so-called orthogonal quantum cryptographic schemes have also
been proposed.
They use only a single basis of communication and,
according to L. Goldenberg, it is possible to use them
to achieve efficiencies
greater than $50\%$ \cite{gold,koashi}. Since they are conceptually
quite different from what we are proposing,
we will not discuss them here.]
By beating this limit, we can better
understand the fundamental limit to the efficiency of quantum cryptography.
Thirdly, our idea is rather general and can be applied to improve the
efficiency of
almost all existing schemes, including the most well-known BB84 scheme.
Finally, the efficiency of quantum cryptography is of
practical importance because it may play a crucial role
in deciding the feasibility of practical quantum cryptographic
systems in any future application.

In Section 2, we review the BB84 scheme.
We introduce our refined error analysis in Section 3 and show that
it generally gives us more power in detecting eavesdropping even
when Alice and Bob choose between the two bases randomly, as in the case of
BB84. In Section 4, we let each of
Alice and Bob choose between the two bases with
biased probabilities $\epsilon$ and $1 - \epsilon$.
We argue in Section~5 that, as far as a biased eavesdropping
attack is concerned, our new scheme
does not lead to a compromise of security if a refined error analysis
is performed.
We also note that our refined analysis is an essential feature
of an improved scheme.
We discuss briefly the subject of
{\it privacy amplification\/} against a biased eavesdropping attack
in Section 6.
The constraint on $\epsilon$ is derived in Section~7.
Section 8 is a collection of concluding remarks:
Firstly, the basic concept of our improved scheme generalizes
trivially to some other
quantum key distribution schemes such as Ekert's scheme and
a scheme based on quantum memories. Secondly, the security issues of our
scheme against other attacks are briefly discussed.
Finally, the history behind this paper is given.

\section{Bennett and Brassard's scheme (BB84)}

In BB84 \cite{BB84}, there are two
participants: the sender, Alice, and the receiver, Bob.
Alice prepares and transmits to
Bob a batch of photons each of which is in one of the four possible polarizations:
horizontal, vertical, 45-degree and 135-degree. Bob measures
the polarizations at the other end. There are two types of
measurements that Bob may perform: He may measure along the
rectilinear basis, thus distinguishing between horizontal
and vertical photons. Alternatively, he may measure
along the diagonal basis, thus distinguishing between
the 45-degree and 135-degree photons.
However, the laws of quantum physics strictly forbid Bob to
distinguish between the four possibilities with certainty.
One way to think of the situation is that the two
polarization bases (rectilinear and diagonal) are complementary
observables. The uncertainty principle of
quantum mechanics dictates that it is impossible
to determine these two observables simultaneously.

Another important feature of quantum mechanics is that
a measurement on an unknown state is generally an irreversible
process that erases the original state.
Suppose Bob chooses to measure the polarization of a photon along
the rectilinear basis. If the photon initially happens to be
diagonally (i.e., 45-degree or 135-degree) polarized, the measurement will
give a {\it random\/} outcome of being either horizontal or vertical.
After the measurement, the photon {\it becomes\/} rectilinearly
polarized (as specified by the measurement outcome)
and completely loses the information on its
initial polarization. For our present discussion, it
suffices to remember the fact that
a measurement along the wrong basis gives a random outcome.

BB84 requires two communications channels between
Alice and Bob. Firstly, there is a public {\it unjammable\/} classical
channel.
i.e., it is assumed that everyone, including the eavesdropper, can listen
to the conversations but cannot change the message.
In practice, an authenticated channel should suffice.
Second, there is a channel for quantum signals. In practice,
the transmission can be done through free air \cite{exp,Buttler,free}
or optical fibres \cite{Hughes96,MZG95,fibers}.
The quantum channel is assumed to be insecure. i.e.,
the eavesdropper is free to manipulate the signals.

In BB84, Alice sends a sequence of photons to Bob.
The protocol consists of several steps:

(1) Alice sends a sequence of photons each in one of the
four polarizations (horizontal, vertical, 45 degrees and 135 degrees)
chosen randomly and independently.

(2) For each photon, Bob chooses the type
of measurement randomly: along either the rectilinear or diagonal bases.

(3) Bob records his measurement bases and the results of the measurements.

(4) Subsequently,
Bob announces his bases (but {\it not\/} the results) through
the public unjammable channel that he shares with Alice.

Notice that it is crucial that Bob announces his basis only after
his measurement. This ensures that during the
transmission of the signals through the quantum channel the eavesdropper Eve
does not know which basis to eavesdrop along.
Otherwise, Eve can avoid detection simply by
measuring along the same basis used by Bob.

(5) Alice tells Bob which of his measurements have been done in the correct bases.

(6) Alice and Bob divide up their polarization data into two classes
depending on whether they have used the same basis.

Notice that Bob should have performed the wrong type of measurements for,
on average, half of the photons. Here, by a wrong type of measurement
we mean that Bob has used a basis different from that of Alice.
For those photons, he gets random outcomes.
Therefore, he throws away those polarization data.
We emphasize that this immediately implies that half of the
data are thrown away and the efficiency of BB84 is bounded by 50\%.

On the other hand, assuming that no eavesdropping has occurred,
all the photons
that are measured by Bob in the correct bases should give the
same polarizations as prepared by Alice. Besides,
Bob can determine those polarizations by his own detectors without
any communications from Alice. Therefore,
those polarization data are a candidate for their raw key.
However, before they proceed any further, it is
crucial that they test for tampering.
For instance, they can use the following
simplified method for estimating the error rate. [Going
through BB84 would give us essentially the same result, namely that all
accepted data
are lumped together to compute a
{\it single\/} error rate.]

(7) Alice and Bob randomly pick a subset of photons from those
that are measured in the correct bases and publicly compare
their polarization data for preparation and measurement.
For those results, they estimate the error rate for the transmission.
Of course, since the polarization data of photons
in this subset have been announced, Alice and Bob must
sacrifice those data to avoid information leakage to Eve.
[This, however, has little effect on the efficiency if the total
number of the transmitted photons is large.]

We assume that Alice and Bob have some idea on the channel characteristics.
If the average error rate $\bar{e}$ turns out to be unreasonably large
(i.e., $\bar{e} \geq e_{\rm max}$ where
$e_{\rm max}$ is the maximal tolerable error rate), then either
substantial eavesdropping has occurred or the channel is somehow
unusually noisy. In both cases, all the data are discarded
and Alice and Bob may re-start the whole procedure again.
Notice that, even then there is no loss in security
because the compromised key is never used to encipher sensitive data.
Indeed, Alice and Bob will derive a key from the data only when
the security of the polarization data is first established.

On the other hand, if the error rate turns out to be reasonably small
(i.e., $\bar{e} < e_{\rm max}$), they go to
the next step.

(8) Reconciliation and privacy amplification: Alice and Bob can
independently convert the polarizations of the remaining
photons into a {\it raw\/} key by, for example,
regarding a horizontal or 45-degree photon as
denoting a `0' and a vertical or 135-degree photon a `1'.

There are still two problems \cite{exp}, namely
noise and leakage of information to Eve. Indeed, the raw key that
Alice has
may differ slightly from that of Bob. It is important for
them to reconcile their differences by performing
error correction (at the cost of throwing away some
polarization data).
We shall skip the details of this reconciliation procedure here.
Now Eve may still have partial information
on the reconciled string between Alice and Bob. A realistic scheme must
include privacy amplification---the distillation
of a shorter but almost perfectly secure key out of a raw key that Eve may
have partial knowledge of.
Privacy amplification schemes
that are secure against single-photon measurements by Eve
have been devised \cite{BBR}.
Let us just mention a useful
result on privacy amplification here.

\subsection{A Result on Privacy Amplification}

Given a string $x$ of length $n$, we say that a {\it deterministic\/}
bit of information about $x$ is the value $f(x)$ of an arbitrary
function $f: \{0,1\}^n \to  \{0,1\}$.
Suppose that there are $n$ bits in a reconciled string $x$
and Eve has at most $l$ deterministic bits of information
about it. The following result is known~\cite{BBR}:
A~{\it hash\/} function
$h$ can be chosen randomly from an appropriate class of
functions $\{ 0,1 \}^n \to \{ 0,1 \}^{n-l-s}$
where $s >0$ such that
the reconciled string $x$ will be mapped into $h(x)$ with
Eve's expected information on $h(x)$ less than $2^{-s}/ \ln 2$ bit.
Alice and Bob can now each compute the value $h(x)$ and keep it as
a secret key for subsequent communication.
More powerful theorems on privacy amplification are given in~\cite{BBCM}
but the above suffices to handle the biased eavesdropping attack that
we analyse here.

\section{Refined Error Analysis}
In the original BB84 scheme, all the accepted data (those
for which Alice and Bob measure along the same basis) are lumped
together to compute a {\it single\/} error rate. In this Section,
we introduce a
refined error analysis. The idea is for Alice and Bob to
divide up the accepted data into two subsets according to
the actual basis (rectilinear or diagonal) used. After that,
a random subset of photons is drawn from each of the two sets.
They then publicly compare their polarization data and
from there estimate the error rate for each basis {\it separately}.
They demand that the run is acceptable if and only if both error
rates are sufficiently small.

In more detail, we keep steps 1) to 5) of BB84 described in Section~2.

6) Recall that each of Alice and Bob
uses the two bases---rectilinear and diagonal---randomly.
Alice and Bob divide up their
polarization data into four cases according to the actual bases used.
They then throw away the two cases
when they have used different bases. The remaining two cases
are kept for further analysis.

7) From the subset where they both use the rectilinear basis, Alice
and Bob randomly pick a fixed number say $m_1$ photons and publicly compare
their polarizations. The number of mismatches $r_1$ tells them
the estimated error rate $e_1 = r_1/m_1$.
Similarly, from the subset where they both use the diagonal basis,
Alice and Bob randomly pick a fixed number
say $m_2$ photons and publicly compare
their polarizations. The number of mismatches $r_2$ gives
the estimated error rate $e_2 = r_2/ m_2$.

Provided that the test samples $m_1$ and $m_2$ are sufficiently large, the estimated
error rates $e_1$ and $e_2$ should be rather
accurate. [The difference between the estimated error rates from
the theoretical error rates can be computed from classical probability theory.]
Now they demand that $e_1, e_2 < e_{\rm max}$ where $e_{\rm max}$
is a prescribed maximal tolerable error rate.
If these two independent constraints are satisfied, they proceed to
step 8). Otherwise, they throw away the polarization data and
re-start the whole procedure from step 1).

8) Reconciliation and privacy amplification: This step is the same as in BB84.

Notice that the two constraints $e_1, e_2 < e_{\rm max}$ are
more stringent than the original naive prescription
$\bar{e} < e_{\rm max}$ in BB84.
To understand this point, consider the following example
of a so-called biased
eavesdropping strategy by Eve.

\subsection{Biased Eavesdropping Strategy}
For each photon, Eve 1) with a probability $p_1$
measures its polarization along the rectilinear basis 
and resends the result of her measurement to Bob;
2) with a probability $p_2$ measures its polarization along the
diagonal basis  and resends the result of her measurement to Bob; and 3)
with a probability $1-p_1 -p_2$, does nothing.
We remark that, by varying the values of $p_1$ and $p_2$,
Eve has a whole class of eavesdropping strategies.
Let us call any of the strategies in this class
a biased eavesdropping attack.

Suppose Alice and Bob know beforehand that the channel
error rate is roughly 2\%. They may decide the maximal
tolerable error rate to be 3\% in BB84. What does this requirement
translate to as constraints on $p_1$ and $p_2$?

Let us compute the error rate for the two bases separately.
When both Alice and Bob use the rectilinear basis,
errors occur only if Eve eavesdrops along the wrong (i.e., diagonal) basis.
This happens with a probability $p_2$. And when Eve uses
the wrong basis,
the polarization of the photon is randomized. Subsequently,
Bob gets an incorrect answer with a probability $1/2$.
Multiplying the two probabilities gives us the error rate $e_1 = p_2/2$
for the rectilinear basis.
A similar argument shows that the error rate for the diagonal basis is
$e_2= p_1/2$.

Now in BB84, all the accepted data are lumped together and a single
error rate is computed. Since the two bases are chosen with
equal probability, the single error rate is
given by
\begin{equation}
\bar{e}=(e_1 + e_2)/2 = (p_1 + p_2) /4 .
\end{equation}
Therefore, the requirement that $\bar{e} < 3 \%$ translates
to $(p_1 + p_2) < 12 \%$.

Now consider our refined error analysis. By computing the
two error rates $e_1$ and $e_2$ separately, it may be natural to
require (for a channel symmetric with respect to
the interchange of the two bases) that $e_1, e_2 < 3 \%$
individually. Now the
two requirements
$e_1, e_2 < 3 \%$ translate into the two constraints
$p_1, p_2 < 6 \%$. They are clearly more stringent than
the single constraint $(p_1 + p_2) < 12 \%$.
For instance, the case when $p_1 =0$ and
$p_2 = 9\%$ violates our improved scheme and will, thus, be
rejected whereas it is acceptable to
BB84.

The usefulness of such a refined error analysis will be
discussed for our efficient scheme in Section~5.

\section{Bias}
The refined error analysis introduced in the last section
is one of the two crucial ingredients
of the improved scheme. This section concerns the second
ingredient---putting a bias in the probabilities of choosing between the
two bases.

Recall the fraction of rejected data of BB84 is at least 50 \%.
This is because in BB84
Alice and Bob choose between the two bases randomly and independently.
Consequently, on average Bob performs a wrong type of measurement
half of the time and, therefore, half of the photons are thrown away
immediately. Here, we propose a simple modification that essentially
doubles the efficiency of BB84. More specifically,
we replace steps 1) and 2) of BB84 described in Section~2 by
the following procedure:

1$'$) Alice and
Bob pick a number $0 < \epsilon \leq 1/2$ whose value is
made public. [Because of the symmetry between the interchange
of the two bases under
$\epsilon \leftrightarrow  1- \epsilon$, there is no
need to consider $\epsilon > 1/2$.]  The value of $\epsilon$ should
be small but {\it non-zero}. The limit $\epsilon \to 0$ is
{\it singular\/} as the scheme is insecure when
$\epsilon =0$.
The constraint on the value $\epsilon$ will be
discussed in Section~7.
Now for each photon Alice chooses between
the two bases, rectilinear and diagonal, with probabilities
$\epsilon$ and $1 - \epsilon$ respectively.

2$'$) Similarly, Bob measures
the polarization of the received photon along the
rectilinear and diagonal bases
with probabilities $\epsilon$ and $1 - \epsilon$ respectively.

We remark that BB84 is a special case of our scheme
when $\epsilon = 1/2$.
In the general case, however,
the bases used by Alice and Bob agree with a probability
$ \epsilon^2 + ( 1 - \epsilon)^2$ which goes to $1$ as $\epsilon$ goes to
zero.
Hence, the efficiency is asymptotically doubled when compared to
BB84.

Notice also that the bias in the probabilities may be produced passively
by an apparatus, for example, an unbalanced beamsplitter. Such a passive
implementation eliminates the need for fast switching
between different polarization bases and is, thus, useful in
experiments.

\section{Refined Error Analysis is Necessary and Sufficient for foiling
the Biased Eavesdropping Attack}
The big question is security.
Naively, one might think that
the knowledge of $\epsilon$ can be exploited by the eavesdropper to
devise a fatal attack.
We remark that this would have been the case if a naive error analysis
(i.e., the estimation of a single error rate)
as prescribed in BB84 had been used. Except for Section~8,
we shall consider only
the biased eavesdropping attack presented in Subsection~3A.
Consider the error rate $e_1$ for the case when both Alice and
Bob use the rectilinear basis. For the biased eavesdropping strategy under
current consideration,
errors occur only if Eve uses the diagonal basis.
This happens with a {\it conditional\/} probability $p_2$.
In this case,
the polarization of the photon is randomized, thus giving an
error rate $e_1 = p_2/2$.
Similarly, errors for the diagonal basis
occur only if Eve is measuring along the
rectilinear basis. This happens with a conditional
probability $p_1$ and when it happens, the photon polarization is
randomized. Hence,
the error rate for the diagonal basis $e_2 = p_1/2$.
Therefore, Alice and Bob
will find that, for the biased eavesdropping attack
of Section~3A, the average error rate
\begin{equation}
\bar{e} = { \epsilon^2 e_1 + ( 1 - \epsilon)^2 e_2  \over
\epsilon^2 +  ( 1 - \epsilon)^2 }
 = { \epsilon^2 p_2 + ( 1 - \epsilon)^2 p_1  \over
2 [ \epsilon^2 +  ( 1 - \epsilon)^2] } .
\end{equation}
Suppose Eve always eavesdrops solely along the diagonal basis (i.e.,
$p_1 =0$ and $p_2 = 1$), then
\begin{equation}
\bar{e} = { \epsilon^2   \over
2 [ \epsilon^2 +  ( 1 - \epsilon)^2] }  \to 0
\end{equation}
as $\epsilon$ tends to $0$.
Hence, with
the original error estimation method
in BB84, Alice and Bob will fail to detect eavesdropping by Eve.
Yet, Eve will have much information about Alice and Bob's raw key
as she is always eavesdropping along the dominant (diagonal) basis.
Hence, a naive error analysis fails miserably.

However, the key point of this paper is the following observation:
The refined error analysis introduced in
Section~3 can make our scheme secure against such a biased eavesdropping
attack.
Recall that in a refined error analysis, the two error rates
are computed {\it separately}. The key observation is that
these two error rates $e_1= p_2/2$ and $e_2= p_1/2$ depend
only on Eve's eavesdropping strategy, but {\it not\/} on
the value of $\epsilon$! This is so because
they are 
{\it conditional\/} probabilities.
This fact is valid not only for the
above biased eavesdropping
strategy, but also for {\it any\/} single-photon eavesdropping
strategy. See Section~8 for a discussion.

More concretely, suppose we
demand that {\it both\/} $e_1$ and $e_2$ are sufficiently
small, say less than $3\%$, we have put a severe constraint on
the amount of information leaked to Eve.
For example, for the biased eavesdropping strategy under current consideration,
the requirements $e_1, e_2 < 3\%$ translate into
$p_1, p_2 < 6\%$. Thus, for those strategies, Eve has at most only
$6\%$ of the information sent by Alice.

\section{Reconciliation and Privacy Amplification}

Now suppose $N$ photons are sent from Alice to Bob.
Most of them (a fraction of about $\epsilon^2 + ( 1 - \epsilon)^2$) will
be accepted data. The reconciliation will lead to some further loss
in efficiency, but not too much.
Suppose the reconciled key $x$ is of length $aN$ where
$a < 1$ is, nonetheless, significantly larger than $1/2$.
Alice and Bob can conservatively estimate the information leakage to Eve to
be $N ( 6 \% + \delta )$ where $\delta$ is a small positive number
that takes into account
the potential error in the statistical estimation of the error rate
and its value can be computed simply from classical probability theory.
The result on privacy amplification in Subsection~2A shows that
Alice and Bob can choose a hash function randomly and publicly from
a class of function to
distill out a key $h(x)$ of length $N ( a - 6 \% - \delta) - s$
where $s > 0$ with the confidence that Eve's expected information is less than
$ 2^{-s} / {\rm ln} 2 $ bits.
Now $h(x)$ is highly secure and can be used a key for subsequent communication.
The key observation is that, for sufficiently small $\epsilon$,
the length of $h(x)$ can be larger than $N/2$, thus
decisively beating the $50\%$ efficiency limit
set by BB84.

This shows that our scheme is perfectly
secure against a biased eavesdropping attack.
The security of our scheme against other attacks is briefly discussed in
Section~8.

\section{Constraint on $\epsilon$.}
Of course, if
$\epsilon$ were actually zero, the improved scheme
would be insecure because
Eve could simply eavesdrop along the diagonal axis.
However, we emphasize that the limit $\epsilon \to 0$ is {\it singular\/}
and that for non-zero $\epsilon$, secure schemes do exist. 
A natural question to ask is:
What is the constraint on $\epsilon$?
The main constraint is that one needs to make sure
that there are enough photons for an accurate estimation of
the two error rates $e_1$ and $e_2$.
Suppose $N$ photons
are transmitted from Alice to Bob. On average, only
$N \epsilon^2$ photons belong to the case where both Alice
and Bob use the rectilinear basis. To estimate $e_1$ reasonably
accurately, one needs to make sure that this number $N \epsilon^2$ is
larger than some fixed number say $m_1$. The key point to
note is that the number
$m_1$ depends on $e_1$ and the desired accuracy of
the estimation but {\it not\/} on $N$. (Indeed,
the number $m_1$ can be computed from classical statistical analysis.)
In summary, one needs:
\begin{eqnarray}
N \epsilon^2 & \geq&  m_1 \nonumber \\
\epsilon &\geq& \sqrt{ m_1 / N} .
\end{eqnarray}
As $N$ tends to infinity, $\epsilon$ can be made to go to zero
but never quite reach it. Notice that
the asymptotic limit $\epsilon \to 0 $ corresponds to $100\%$
efficiency.
In conclusion, the improved
scheme is asymptotically the most efficient scheme that one
can possibly devise.

\section{Concluding Remarks}
In BB84, each of Alice and Bob chooses between the two bases (rectilinear and
diagonal) with equal probability. Consequently, Bob's measurement
basis differs from that of Alice's half of the time. For this reason,
half of the polarization data are useless and are thus thrown away immediately. 
We have presented a simple modification that can essentially double the
efficiency of BB84. There are two important ingredients in this modification.
The first ingredient is for each of Alice and Bob to assign significantly
different probabilities (say $\epsilon$ and $1 - \epsilon $ respectively where
$\epsilon$ is small but non-zero) to the two polarization bases (rectilinear
and diagonal respectively). Consequently, they are much more likely to
use the same basis. This decisively enhances efficiency.

However, an eavesdropper may try to break such a scheme by
eavesdropping mainly along the predominant basis.
To make the scheme secure against such a biased eavesdropping attack,
it is crucial to have the second
ingredient---a refined error analysis---in place.
The idea is the following. Instead of lumping all the accepted polarization
data into one set and computing a {\it single\/} error rate (as in BB84), we
divide up the data into various subsets according to the actual polarization
bases used by Alice and Bob. In particular, the {\it two\/} error rates for
the cases 1) when both Alice and Bob use the rectilinear basis and
2) when both Alice and Bob use the diagonal basis, are computed separately.
It is only when both error rates are small that they accept the security of
the transmission. We have demonstrated that this refined analysis is necessary
and sufficient in guaranteeing the security of our improved scheme
against a biased eavesdropping attack.

We remark that our idea of efficient schemes of quantum key distribution
applies also to other schemes such as
Ekert's scheme \cite{Ekert} and Biham, Huttner and Mor's
scheme \cite{Mor} which is
based on quantum memories.

As a side remark, Alice and Bob may use different biases
in their choices of probabilities. In other words,
our idea still works if Alice chooses between
the two bases with probabilities $\epsilon$ and  $1- \epsilon$
and Bob chooses with probabilities $\epsilon'$ and  $1- \epsilon'$
where $\epsilon \not= \epsilon'$.

So far our discussion on security has been restricted to a
biased eavesdropping attack.  What about its security against
other attacks?  Indeed, it is a highly non-trivial problem to work out
security even for the standard BB84 scheme, and even if we restrict
the eavesdropper to attacking Alice's photons one by one.
We certainly do not claim
to have fully worked out the security of our new scheme.  However,
in the near future, it is plausible that single-photon-measurement attacks by
Eve will be the only realistic class of attacks.  It~would be interesting
to prove that our scheme is at least as secure as BB84 against any
such restricted attack.  This hope is reasonable because any
single-photon-measurement eavesdropping
strategy gives characteristic error rates $e_1$ and
$e_2$ {\it independent\/} of the value of $\epsilon$.
This is so because they are {\it conditional\/} probabilities.
Consequently, it is intuitively plausible that
Eve cannot exploit her knowledge of $\epsilon$
to avoid detection of her tampering attempt.

Finally, a piece of history on this paper.
Apparently, the possibility of having more efficient quantum key distribution
schemes was first raised by one of us (M. Ardehali)
in an unpublished manuscript \cite{eff}.
Unfortunately,
the crucial importance of a refined error analysis was not recognized.
As pointed out by G. Brassard,
the security of that scheme remained unproven.
The use of a refined error analysis
was first discussed by
Barnett and Phoenix\cite{Barnett} for {\it rejected\/} data.
Two of us (H.-K. Lo and H. F. Chau), however, noted \cite{patent} the important
fact that when a refined error analysis
is applied to {\it accepted\/} data, an improved scheme can be made
secure.

We thank Gilles Brassard for helpful discussions and suggestions.

\par\bigskip\noindent
{\it Notes Added}: An entanglement-based scheme with an
efficiency greater than
$50\%$ has also been discussed in a recent preprint by two of us \cite{qkd}.

\end{document}